\documentclass{PoS}
\usepackage{epsfig}

\title{Electric polarizability of the neutron in dynamical quark ensembles}

\ShortTitle{Electric polarizability of the neutron in dynamical quark ensembles}

\author{\speaker{Michael Engelhardt}\\
        Department of Physics, New Mexico State University\\
        E-mail: \email{engel@nmsu.edu}}

\abstract{The background field method for measuring the electric
polarizability of the neutron is adapted to the dynamical quark case,
resulting in the calculation of (certain space-time integrals over)
three- and four-point functions. Particular care is taken to disentangle
polarizability effects from the effects of subjecting the neutron
to a constant background gauge field; such a field is not a pure gauge on
a finite lattice and engenders a mass shift of its own. At a pion mass
of $m_{\pi } =759\, \mbox{MeV} $, a small, slightly negative electric
polarizability is found for the neutron.}

\FullConference{The XXV International Symposium on Lattice Field Theory\\
                 July 30 - August 4, 2007\\
                 Regensburg, Germany}

\begin{document}

\section{Introduction}
The response of hadrons to external electromagnetic fields is characterized
by their polarizabilities. Polarizabilities manifest themselves, e.g., in
soft Compton scattering; the incoming electromagnetic wave distorts the
hadron, and this distortion in turn leads to a modified scattered wave.
Accordingly, polarizabilities appear as effects of second order in the
external field in the effective hadron Hamiltonian. In a low-energy
expansion of that Hamiltonian, the leading term embodies the hadron
mass shift in constant electric and magnetic fields $E$ and $B$,
\begin{equation}
H_{\mbox{\scriptsize eff} } =
\frac{1}{2} \left( \alpha E^2 + \beta B^2 \right)
\label{poldef}
\end{equation}
with the static electric and magnetic polarizabilities $\alpha $ and
$\beta $, respectively. Carrying the low-energy expansion to subsequent
orders, dependences on particle spin and electromagnetic field gradients
appear \cite{polariz}.

The work presented here is concerned specifically with the static electric
polarizability $\alpha $ of the neutron. There are two chief new elements
to this investigation. On the one hand, dynamical quark ensembles are used;
on the other hand, the role of constant background gauge fields is
recognized and their effects are properly taken into account. As far as
the former issue is concerned, it should be noted that, when introducing
an explicit external electromagnetic field to compute polarizabilities on
the lattice, the transition from quenched to dynamical quark ensembles
entails more than just the usual increased computational cost of generating
the ensemble. In the quenched approximation, the backreaction of the quark
fields on the gluonic degrees of freedom is truncated; thus, the external
electromagnetic field, which in turn only couples to the quark fields,
does not influence the gauge ensemble. As a result, one can generate the
ensemble in the absence of the background field and introduce the latter
a posteriori by a suitable modification of the link variables. By contrast,
in the dynamical quark case, at first sight, one would need to recompute
a new gauge ensemble for every specific external field considered, a
prohibitively expensive approach if directly applied. Previous quenched
investigations \cite{fiebpol,wilcoxe,wilcoxnew} were thus able to take
advantage of considerable simplifications not available for dynamical
quarks. The dynamical quark case necessitates significant modifications
of the computational scheme in order to become tractable,
cf.~section \ref{pertexp} below.

\begin{figure}[h]
\centerline{\epsfig{file=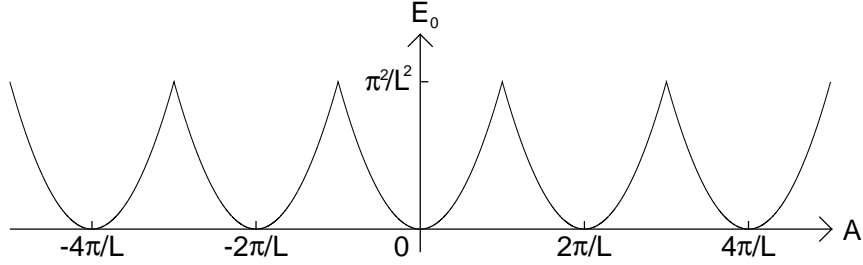,width=4cm,angle=-90} }
\caption{Ground state energy of a charged particle on a circle of length
$L$ with periodic boundary conditions, as a function of the constant
gauge field $A$ to which it is subjected, cf.~main text.}
\label{spflow}
\end{figure}

Concerning the issue of constant background gauge fields, consider
introducing a constant electric field in the 3-direction via the gauge
field
\begin{equation}
A_3 (t) = E(t-t_0 ) \equiv A+Et \ .
\label{bgfield}
\end{equation}
For given electric field $E$, one still has a continuum of choices for
$t_0 $ or, equivalently, $A$. In a setting in which the spatial directions
are infinite, this is merely a gauge freedom; $A$ can be arbitrarily
shifted by gauge transformations. However, if the spatial directions are
finite, such as in a lattice calculation, this gauge freedom is reduced by
the need to preserve the boundary conditions; as a consequence, physics in
general depends on the value of $A$. As a simple example, consider a
charged particle on a circle of length $L$ with periodic boundary
conditions in a constant gauge field $A$. The Hamiltonian is $H=(p-A)^2 $
and the energy levels are $E_n = (2\pi n/L -A)^2 $, where $n$ is any whole
number. In particular, the ground state energy as a function of $A$ is
plotted in Fig.~\ref{spflow}.
Only a discrete symmetry under shifts $A \rightarrow A+2\pi /L$ remains;
$A$ can be interpreted as a Bloch momentum of the charged particle.

Similarly, the mass shift of a neutron in the external field (\ref{bgfield})
is a function of both the parameters $E$ and $A$. The dependence on $A$ is
a dominant effect which must disentangled from the polarizability in presence
of the electric field $E$.

\section{Perturbative expansion}
\label{pertexp}
\begin{figure}[h]
\centerline{\epsfig{file=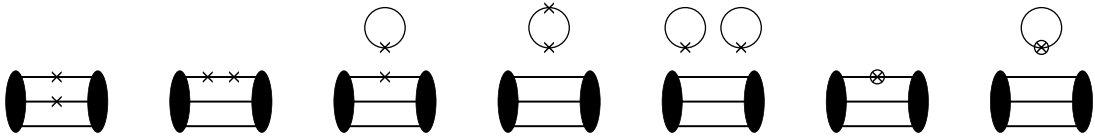,width=15cm} }
\vspace{-0.5cm}
\caption{Diagrams of second order in the external field resulting from
inserting the electromagnetic vertices generated by the interaction
term $S_{ext} $ into a neutron two-point function. Crosses denote
vertices linear in the background field, circled crosses vertices
quadratic in the external field.}
\label{fig4pt}
\end{figure}

As mentioned above, in the presence of dynamical quarks, the gauge ensemble
depends on the specific external electromagnetic field under consideration.
However, it would be prohibitively expensive to recompute the ensemble for
each background field. Relief is provided by the fact that it is not
necessary to be able to evaluate the neutron mass shift in arbitrary
backgrounds; after all, the electric polarizability is given specifically
by the second Taylor coefficient of that mass shift as a function of the
electric field, cf.~(\ref{poldef}). To evaluate a Taylor coefficient, it
is sufficient to consider only infinitesimal external fields, and in this
case, one can correspondingly expand the neutron two-point function
determining the mass shift. Introducing the external field $A_3 $,
cf.~(\ref{bgfield}), into the link variables as an additional $U(1)$ phase,
\begin{equation}
U_3 \longrightarrow
\exp \left( i\ \int dx_3 \cdot A_3 \right) \cdot U_3
= \left( 1 + iaA_3 - a^2 A_3^2 /2 + \ldots \right) \cdot U_3 \ ,
\end{equation}
one obtains an action
\begin{equation}
S = S_0 + S_{ext}
\label{totact}
\end{equation}
composed of the QCD action in the absence of the external field, $S_0 $,
and the coupling to the external field
\begin{eqnarray}
S_{ext} &=& z_V \frac{1}{2} \Sigma_{x} \ \bar{\psi } (x) \left[
(iaA_3 - a^2 A_3^2 /2) \cdot U_3 (x) \cdot (-1+\gamma_{3} )
\cdot \psi (x+e_3 ) ) \right. \label{sext} \\
& & \hspace{3cm} \left.
+(iaA_3 + a^2 A_3^2 /2) \cdot U^{\dagger }_{3} (x-e_3 ) \cdot
(1+\gamma_{3} ) \cdot \psi (x-e_3 ) ) \right] \ ,
\nonumber
\end{eqnarray}
which has been written here specifically for Wilson-type four-dimensional
quark fields. The particular computational scheme adopted in the following
uses domain wall fermions in the valence sector; however, quark propagators
are evaluated starting from sources which only have support on the
four-dimensional domain walls, and they are projected back onto four
dimensions at the sink. This is a compromise due to hardware constraints
on the storage of propagators; correspondingly, the coupling to the
external field is effected via the four-dimensional projected fields
$\psi $, as written in (\ref{sext}). Concomitantly, it is necessary to
allow for the renormalization factor $z_V $ on quark bilinears to
compensate for the effect of the projection. This factor was determined
in \cite{elpol} by evaluating the number of valence quarks in the neutron,
yielding $z_V =1.12\pm 0.12$. Inserting the action (\ref{totact}) into the
neutron two-point function, one has
\begin{eqnarray}
\left\langle N_{\beta } (y) \bar{N}_{\alpha } (x) \right\rangle &=&
\int [DU] [D\psi ] [D\bar{\psi } ] \ \exp (-S_0 ) \
\left( 1 - S_{ext} + S_{ext}^{2} /2 + \ldots \right)
N_{\beta } (y) \bar{N}_{\alpha } (x) \ /
\nonumber \\
& & \hspace{2cm} \int [DU] [D\psi ] [D\bar{\psi } ] \ \exp
(-S_0 ) \ \left( 1 - S_{ext} + S_{ext}^{2} /2 + \ldots \right)
\label{twopt}
\end{eqnarray}
To obtain specifically the part of the two-point function which is
quadratic in the external field, one therefore must calculate neutron
two-point functions with either two additional insertions of the vertex
linear in the external field, cf.~(\ref{sext}), or one additional
insertion of the vertex quadratic in the external field, cf.~(\ref{sext}),
as displayed in Fig.~\ref{fig4pt}.

\noindent
The space-time positions of the insertions are to be integrated over
according to (\ref{sext}); thus, one is effectively calculating certain
space-time integrals over three- and four-point functions. These functions
are evaluated in the unperturbed ensemble generated by the action $S_0 $.
The two right-most diagrams involving the vertex quadratic in the external
field act as contact terms renormalizing the second and fourth diagrams
from the left, respectively. As usual, the denominator of (\ref{twopt})
effects the subtraction of statistically disconnected pieces from diagrams
containing disconnected quark loops.

\section{Extracting the mass shift}
Focusing on the zero-momentum, unpolarized component of the neutron
two-point function,
\begin{equation}
G(p=0,t) = \Sigma_{\vec{y} } \ \mbox{Tr} \left( \frac{1+\gamma_{0} }{2}
\left\langle N (y) \bar{N} (x) \right\rangle \right)
\ \ \stackrel{t\rightarrow \infty }{\longrightarrow } \ \
W \ \exp (-m \, t)
\end{equation}
one observes an exponential decay in time at a rate governed by the neutron
mass $m$; $W$ characterizes the overlap between neutron sources/sinks and
the neutron ground state wave function. In the setting discussed here, some
care must be exercised in interpreting the two-point function \cite{elpol}.
The Hamiltonian in the presence of the external field (\ref{bgfield}) is
time-dependent, since a translation in time is equivalent to a shift
in the constant gauge field $A$, cf.~(\ref{bgfield}); on the other hand,
a change in $A$ modifies the physical spectrum. For infinitesimal 
external fields, this time dependence can be treated adiabatically, i.e.,
the exponential decay of the two-point function can still be interpreted
in terms of the neutron mass, which, however, will continuously adjust
as the external field changes. Expanding the overlap $W$ and the neutron
mass $m$ in terms of the external field parameters $A$ and $E$,
\begin{eqnarray}
W &=& W_0 + W^{(1)} (A,E) + W^{(2)} (A,E) +\ldots \\
m &=& m_0 + m^{(2)} (A,E) +\ldots
\label{mexpan}
\end{eqnarray}
where the superscripts denote the order in the external parameters, the
second-order part of\linebreak $G(p=0,t)$ behaves as
\begin{equation}
G^{(2)} (p=0,t) \ \ \stackrel{t\rightarrow \infty }{\longrightarrow } \ \
W_0 \ \exp (-m_0 \, t) \left( \frac{W^{(2)} (A,E)}{W_0 }
-m^{(2)} (A,E) t \right) \ .
\label{correl2}
\end{equation}
Note that, in (\ref{mexpan}), use has been made of the fact that there can
be no first-order term in the neutron mass in the absence of a nontrivial
$\theta $-angle. Thus, dividing (\ref{correl2}) by the unperturbed
correlator $W_0 \ \exp (-m_0 \, t)$ yields a ratio whose slope as a
function of time, $S(A,E)$, corresponds, up to a minus sign, to the mass
shift of the neutron in the presence of the external electromagnetic field,
\begin{equation}
-S(A,E) = m^{(2)} (A,E) \equiv m_2^{AA} A^2 +m_2^{AE} AE +m_2^{EE} E^2 \ .
\end{equation}
Here, again, care must be taken in interpreting the results. As mentioned
above, the spectrum of the system adjusts adiabatically to the external
field, i.e., $m^{(2)}$ and $W^{(2)}$ themselves in general are
time-dependent. This complicates the interpretation of $S$ in terms of
the mass shift $m^{(2)} $. However, at the particular point where the
physical spectrum is stationary in $A$, the time dependence of $m^{(2)}$
and $W^{(2)}$ is of at least quadratic order. This is due to the
equivalence between shifts in $A$ and shifts in time, cf.~(\ref{bgfield});
stationarity in $A$ implies stationarity in time. Thus, at the stationary
point only, which in practice is identified by seeking out the extremum of
the slope $S(A,E)$ as a function of $A$, one can indeed directly interpret
that slope as giving $m^{(2)} =-S$.

In turn, as discussed in more detail in \cite{elpol}, the part of the
neutron mass shift due specifically to its electric polarizability can be
isolated from the combined dependence of $m^{(2)} $ on $E$ and $A$ by
seeking out the extremum of $m^{(2)} (A,E)$ (or, equivalently, $S(A,E)$)
as a function of $A$. The mass shift obtained at the extremum then yields
the electric polarizability via
\begin{equation}
\alpha = -\left. \frac{2}{E^2 } m^{(2)}
\right|_{\mbox{\scriptsize extremum} } \ ,
\label{mtoa}
\end{equation}
cf.~(\ref{poldef}). To carry out this program in practice, one thus needs
measurements at three different values of $A$ (or, equivalently, $t_0 $)
in (\ref{bgfield}), while $E$ is fixed. This then defines a parabola in
$t_0 $, permitting the extraction of the extremum as a function of $t_0 $.

\section{Measurements}
The program described above was put into practice using 99 configurations
from one of the dynamical 2+1 flavor Asqtad quark ensembles provided by
the MILC collaboration \cite{milc}. Specifically, the $SU(3)$
flavor-symmetric ensemble with quark masses $am_u =am_d =am_s =0.05$ was
utilized, where $a=0.124\, \mbox{fm} $ denotes the lattice spacing. This
choice of quark masses corresponds to a pion mass of
$m_{\pi } =759\, \mbox{MeV} $. On the other hand, as already mentioned
in section \ref{pertexp}, in the valence sector, domain wall quarks were
used; this hybrid scheme has been previously employed in
\cite{gapaper,gpdpaper}, where also details concerning parameter
tuning in this approach can be found.

Table~\ref{sloptab} displays the measurements obtained for the slope
$S(A,E)$, together with a determination of its extremum as a function
of $A$, which yields (up to a minus sign) the mass shift due to the
electric polarizability of the neutron. In all measurements, the neutron
source was placed at $t=0$ and the slope $S(A,E)$ was extracted by fitting
in the range $4a\leq t\leq 10a$. The disconnected traces were evaluated
using stochastic estimation with 120 complex $Z(2)$ stochastic sources
(240 for the case $t_0 =-10a$, which displayed particularly strong
statistical fluctuations).
\begin{table}
\begin{center}
\begin{tabular}{|c||c|c|}
\hline
& connected & all \\
& diagrams & diagrams \\
\hline \hline
\parbox{0.001cm}{\vspace{1cm} }
\parbox{2.2cm}{\centerline{ $S/(a^3 E^2 )$ }
\centerline{ $t_0 = -10a$ } } & $0.46(18)$ & $0.26(26)$ \\
\hline
\parbox{0.001cm}{\vspace{1cm} }
\parbox{2.2cm}{\centerline{ $S/(a^3 E^2 )$ }
\centerline{ $t_0 =0$ } } & $0.000(16)$ & $-0.033(43)$ \\
\hline
\parbox{0.001cm}{\vspace{1cm} }
\parbox{2.2cm}{\centerline{ $S/(a^3 E^2 )$ }
\centerline{ $t_0 =6a$ } } & $-0.017(3)$ & $-0.037(25)$ \\
\hline \hline
\parbox{0.001cm}{\vspace{1cm} }
\parbox{2.2cm}{\centerline{ $-m^{(2)} /(a^3 E^2 )$ }
\centerline{ (extremum) } } & $-0.034(6)$ & $-0.052(24)$ \\
\hline
\end{tabular}
\vspace{0.2cm}
\caption{Measurements of the slope $S(A,E)$, together with a determination
of its extremum as a function of $A$. The electric field $E$ is cast in
Gaussian units. Errors were obtained by the jackknife method, including
the fluctuations of the renormalization constant $z_V $.}
\vspace{-0.4cm}
\label{sloptab}
\end{center}
\end{table}
Using (\ref{mtoa}) to obtain the electric polarizability of the neutron
from the last line of Table~\ref{sloptab}, one finally arrives at
\begin{equation}
\alpha = (-2.0\pm 0.9)\cdot 10^{-4} \, \mbox{fm}^3
\label{aresult}
\end{equation}
at a pion mass of $m_{\pi } =759\, \mbox{MeV} $.

\section{Discussion}
\begin{figure}[h]
\vspace{-0.3cm}
\centerline{\epsfig{file=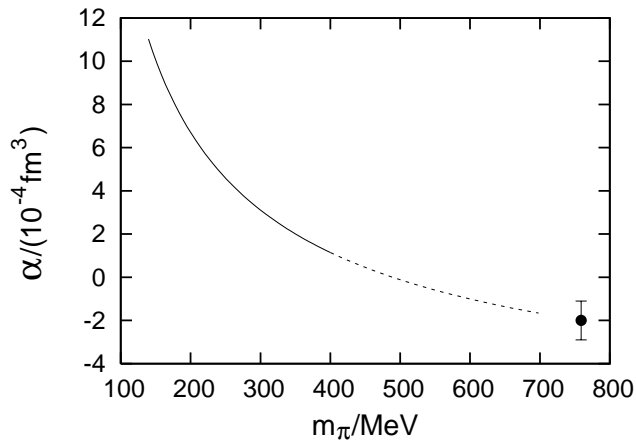,width=6cm,angle=-90} }
\caption{Comparison of the lattice measurement of the electric
polarizability $\alpha $ of the neutron obtained in this work with the pion
mass dependence expected from the ``Small Scale Expansion'' approach
\cite{kambor,hgrie}. This Chiral Effective Theory is expected to be
accurate up to pion masses of about $300\, - 400\, \mbox{MeV} $
(solid line); the dashed line is obtained if one naively continues
the Chiral Effective Theory expression to higher pion masses, beyond
its realm of controlled quantitative applicability.}
\label{chet}
\end{figure}

Compared to the experimental value reported by the Particle Data Group
\cite{pdgroup}, namely,
$\alpha = (11.6\pm 1.5) \cdot 10^{-4} \, \mbox{fm}^3 $,
the result (\ref{aresult}) implies a strong variation with pion mass.
Such a variation is indeed expected from Chiral Effective Theory
\cite{kambor,hgrie,detmold}. Fig.~\ref{chet} displays the result
(\ref{aresult}) in relation to the pion mass dependence expected from
the ``Small Scale Expansion'' approach \cite{kambor,hgrie}, a systematic
extension of leading-one-loop Heavy Baryon Chiral Perturbation Theory by
explicit $\Delta $ degrees of freedom.
While at a pion mass of $m_{\pi } =759\, \mbox{MeV} $, Chiral Effective
Theory cannot be taken as more than a qualitative hint, the comparison
displayed in Fig.~\ref{chet} is tantalizing and calls for a push towards
lower pion masses, where a quantitative connection with Chiral Effective
Theory can be made. In view of the computational effort which was required
for the present dynamical quark calculation at $m_{\pi } =759\, \mbox{MeV} $,
such further progress is feasible, but will require significant resources
by current standards.

\acknowledgments
\vspace{-0.2cm}
This investigation benefited from various exchanges with R.~Brower,
M.~Burkardt, W.~Detmold, R.~Edwards, H.~Grie\ss hammer, D.~B.~Kaplan,
J.~Negele, K.~Orginos, J.~Osborn, D.~Renner, M.~Savage, D.~Toussaint and
W.~Wilcox. Computations were carried out using resources provided by the
U.S.~DOE through the USQCD project at Jefferson Lab. This work is
supported by the U.S.~DOE under grant number DE-FG03-95ER40965.
\vspace{-0.1cm}


\end{document}